\documentclass[aps,twocolumn,floatfix]{revtex4}
\usepackage{epsfig}

\newcommand{\be}{\begin{equation}}
\newcommand{\ee}{\end{equation}}
\newcommand{\ba}{\begin{array}}
\newcommand{\ea}{\end{array}}

\begin{document}
\title{Are there really phase transitions in 1-d heat conduction models?}
\author{Lei Yang and Peter Grassberger}
\date{\today}
\affiliation{{John-von-Neumann Institute for Computing, Forschungszentrum J\"{u}lich},\\
D-52425 J\"{u}lich, Germany}

\begin{abstract}
Recently, it has been claimed (O. V. Gendelman and A. V. Savin, Phys. Rev.
Lett. {\bf 84}, 2381 (2000); A.V.Savin and O.V.Gendelman, arXiv:
cond-mat/0204631 (2002)) that two nonlinear classical 1-d lattice models
show transitions, at finite temperatures, where the heat conduction changes
from being finite to being infinite. These are the well known
Frenkel-Kontorova (FK) model and a model for coupled rotators. For the FK
model we give strong theoretical arguments why such a phase transition is
not to be expected. For both models we show numerically that the effects
observed by Gendelman {\it et al.} are not true phase transitions but are
rather the expected cross-overs associated to the conductivity divergence as 
$T\to 0$ and (for the FK model) $T\to\infty$.
\end{abstract}
\maketitle

Heat conduction in classical one-dimensional lattices has recently been
investigated by many authors (see \cite{lepri} for a recent review). A large
class of 1-d systems can be described by the general Hamiltonian 
\begin{equation}
H=\sum_{i=1}^N\left( \frac{p_i^2}{2m}+U(q_{i+1}-q_i)+V(q_i)\right)
\end{equation}
where $N$ is the system size, $p_i$ is the momentum of the $i$th particle, $
q_i$ is its displacement from the equilibrium position, $U$ is the internal
potential, and $V$ is an external potential. In the following we shall always
use $m=1$, without loss of generality. Real isolated systems must have 
$V=0$, since any external potential would itself involve a ``scaffolding''
which is not rigid and would thus also contribute to heat conduction. Thus,
for real systems the support of the external potential must be included in
the description of the system, and $V$ has to be replaced by a contribution
to the internal potential $U$. Nevertheless, we shall in the following keep
the ansatz (1), understanding that we are calculating only part of the
complete heat conduction when $V\neq 0$. In the following we shall also
assume that the system has no frozen disorder.

It is well known that heat conduction is infinite, if all potentials are
harmonic \cite{harmonic} or if the system is integrable. In this case
phonons resp. solitons are not scattered. Thus they propagate ballistically,
given a constant heat flux $J$ (independent of $N$) when a finite constant
temperature difference $\Delta T$ is applied to the two ends of a chain of
length $N$. Thus formally, the conductivity $\kappa =JN/\Delta T$ is
proportional to $N$. For non-integrable models with an acoustic phonon
branch, i.e. without an external potential, one expects ballistics transport
in the infinite wave length limit (since phonon scattering in general
decreases with energy), which leads to a power behavior $\kappa \sim
N^\alpha $ with $0<\alpha <1$ \cite{foot1}. This is the case, e.g., for the
Fermi-Pasta-Ulam (FPU) model \cite{FPU_1d} and for the diatomic Toda lattice 
\cite{Toda_1d}.

An exception to this seems to happen for coupled rotators \cite
{R_livi,R_savin}, 
\begin{equation}
U(q_{i+1}-q_i)=-\cos (q_{i+1}-q_i)\;,\qquad V(q_i)=0\;,
\end{equation}
which seem to have finite $\kappa $ \cite{R_livi}. This is explained 
\cite{R2_savin} by the fact that single highly excited rotators essentially
decouple from neighboring rotators, acting thus as barriers for the
propagation of any phonons, even soft ones.

Soft acoustic phonons are essentially Goldstone modes due to Galilei
invariance. Finite heat conduction (in $d=1$) is therefore expected when $
V\neq 0$, since then translation invariance is broken and an acoustic branch
does not exist in the phonon spectrum. This is e.g. the case for the
Frenkel-Kontorova (FK) model 
\begin{equation}
U(q_{i+1}-q_i)={\frac 12}(q_{i+1}-q_i)^2\;,\quad V(q_i)=-\varepsilon \cos
q_i\;.
\end{equation}
(notice that this is the {\it commensurate} FK model, where the harmonic part 
of the potential leads to the same particle distance in the ground state as 
the cosine potential; we shall only discuss this case in the following). In 
this model, all phonons have a finite mean free path, bounded from above by 
a finite constant which is independent of the wave number $k$, but which
diverges for $T\to 0$ and for $T\to \infty $. The latter follows from the
fact that $V$ given by Eq.(3) effectively becomes negligible when $T\to
\infty $, and effectively becomes a sum of harmonic potentials when $T\to 0$
. Thus there is no ballistic transport, and no obvious mechanism which could
lead to an infinite conductivity for any finite $T$, while one expects $
\kappa $ to diverge when $T\to 0$ or $T\to \infty $ \cite{FK_bishop}.

It was thus very surprising when Savin and Gendelman \cite{FK_savin} claimed
to have clear evidence for phase transitions in the FK model, at which the
conductivity changed from finite to infinite. They claimed that $\kappa$ 
is finite only in an interval $T_{c1} < T < T_{c2}$, with $T_{c1}$ and $
T_{c2}$ dependent on $\varepsilon$, while $\kappa = \infty$ outside this
interval.

Indeed, the same authors had also claimed that there is a phase transition
in the rotator model \cite{R_savin,R2_savin}. There, the density of highly
excited rotators should of course go to zero for $T\to 0$. Thus one expects
that soft phonons exist in this limit, and $\kappa \to \infty$ for $T\to 0$.
Instead of this, it was claimed in \cite{R_savin,R2_savin} that $
\kappa=\infty$ in an entire interval $0\le T \le T_c$.

In the present paper we want to test these claims by performing simulations
on larger lattices and with higher precision than in \cite{R_savin,R2_savin}.

In order to mimic the simulations of \cite{FK_savin,R_savin,R2_savin} as
close as possible, we also used Langevin thermostats (we do not agree with
these authors that Nos\'{e}-Hoover thermostats would be unsuitable, but we
just don't want any discussion about this point). More precisely, we
simulated a chain of $N_0+N+N_0$ oscillators. The central $N$ oscillators
follow their Hamiltonian equations of motion, while the outer $2N_0$ ones
satisfy $\ddot{q}_n=-\frac{\partial \phi }{\partial q_n}-\gamma \dot{q}
_n+\xi _n$ and with $\phi =U+V$ being the total potential, $\xi _n$ being
white Gaussian noises, $\langle \xi _n(t)\xi _k(t')\rangle =2\gamma
T_n\delta _{nk}\delta (t-t')$ and with $T_n=T_{{\rm high}}$ for $-N_0\le 
n\le 0$ and $T_n=T_{{\rm low}}$ for $N\leq n<N+N_0$. We used $N_0=40$ 
and $\gamma =0.1$, as in \cite{FK_savin,R_savin,R2_savin}. The temperature 
difference $\Delta T\equiv T_{{\rm high}}-T_{{\rm low}}$ was chosen
between 10\% and 20\%.

For the integration we used a simple leap frog \cite{LeapFrog}. On the one 
hand this is symplectic and thus more suited for the central region than, say, 
a Runge-Kutta integrator. On the other hand it should be more robust than
higher order symplectic integrators in the boundary regions which are not
Hamiltonian. Step size was 0.05, and total integration times were typically $
\simeq 10^7-10^8$ units (i.e. $10^8-10^9$ steps), with some runs going up 
to $5\times 10^8$ units. We checked that this was sufficient to reach a steady 
state and that the time-averaged heat flux $J$ was independent of the site. 

\begin{figure}
\psfig{file=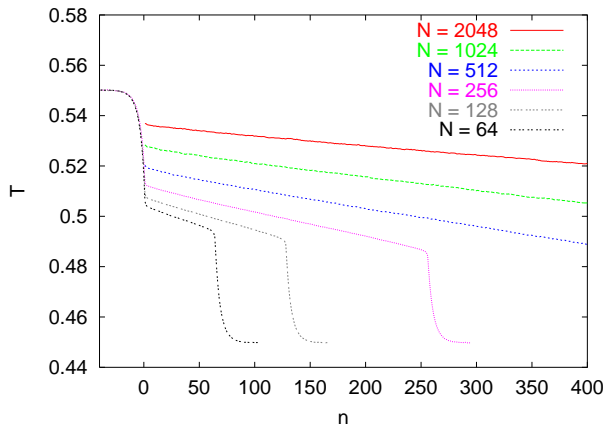,width=5.8cm, angle=270}
\caption{Temperature profiles for the FK model with $\varepsilon=1$ and
$T_{{\rm high}}=0.55$, $T_{{\rm low}}=0.45$. The lengths of the central parts
of the chains are 64, 128, 256, 512, 1024, and 2048. While the gradients in
the central region are roughly constant, most of the temperature variation
happens for short chains in the thermostated boundaries. }
\end{figure}

\begin{figure}
\vglue -4.mm
\psfig{file=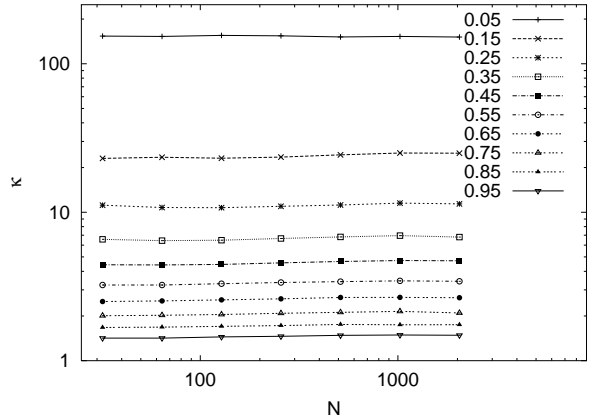,width=5.8cm, angle=270}
\caption{Heat conductivity versus system size for the $\varphi^4$
model. Each curve corresponds to a constant value of $\beta$, while $T=2$
is held fixed. Statistical and integration errors are less than the
symbols. The continuous lines are only drawn for guiding the eye.}
\vglue -2.mm
\end{figure}

We verified also that the temperature profiles were roughly linear in the 
central region $0\le n \le N$, but we could not verify the absence of 
temperature jumps at its boundaries claimed in \cite{FK_savin,R_savin,R2_savin}.
More precisely, such jumps were absent only for very large lattices and small
conductivities, i.e. if the heat flux was small. Otherwise, for small lattices
and/or large conductivities, there were very large jumps, mostly located in 
the boundary regions $-40 < n < 0$ and $N < n < N+40$ (see Fig.~1). Thus while 
the profile in the central region was essentially linear (in contrast to 
simulations with Nose-Hoover thermostats coupled to single particles, as e.g. 
in \cite{FPU_yg,FPU_1d}, it would be very wrong to estimate the conductivity
simply by dividing the flux by the nominal imposed $\Delta T$. It seems that 
this was done in several cases in \cite{FK_savin,R_savin,R2_savin}, which explains 
some -- but not all -- of the differences between their results and those of 
the present paper. In other cases the authors of \cite{FK_savin,R_savin,R2_savin}
must have taken into account boundary jumps, otherwise their results would 
disagree much more with ours than they actually do. Unless 
otherwise said, we will estimate $\kappa$ by dividing the flux by the 
temperature drop over the inner half of the central region.

Notice that a similar behavior was found also for the rotator model and 
for the discrete $\phi^4$ model. The latter is given by the Hamiltonian
\cite{Fai4 zhao}
\be
  H=\frac 12 \sum_i\left( p_i^2+(q_{i+1}-q_i)^2+\alpha q_i^2+\frac 12\beta
q_i^4\right) .
\ee
Conductivities for the $\phi^4$ model with $T=2$ and $\beta =1-\alpha $ are 
shown in Fig.~2 for various values of $\beta \in [0,1]$. The $\phi^4$ model
is a prototype model with finite conductivities. Indeed we see that 
all measured values of $\kappa$ are not only finite but are independent of 
$N$. This would not have been the case if we had not taken the temperature
jumps into account and would have used the nominal value of $\Delta T$ when 
estimating $\kappa$ from $J/\Delta T$. The values of $\kappa$ shown in 
Fig.~2 are in very good agreement with those of \cite{Fai4 zhao}.

{\bf Frenkel-Kontorova model}: In all simulations, $N$ ranged from 32 to
2048. We made simulations for $\varepsilon = 1.0, 3.0$ and 10.0. According 
to \cite{FK_savin}, $T_{c1}$ is only weakly dependent on $\varepsilon$:
$T_{c1}=2.6, 2.3$, and $2.0$ for $\varepsilon = 1.0, 3.0, 10.0$. On the 
other hand, $T_{c2}$ should strongly increase with $\varepsilon$,
$T_{c2} = 3.3, 15$, and 150 for the above three values.

\begin{figure}
\psfig{file=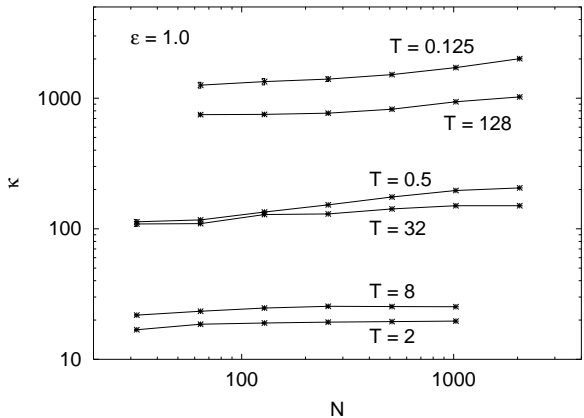,width=5.8cm, angle=270}
\caption{Heat conductivity in the FK model versus system size, for 
$\varepsilon = 1.0$. Each curve corresponds to a constant average temperature.}
\end{figure}

\begin{figure}
\psfig{file=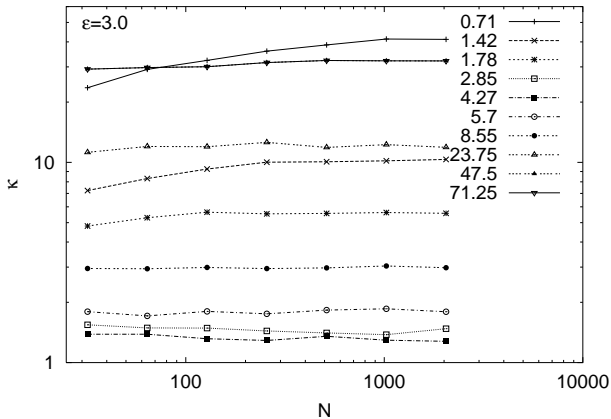,width=5.8cm, angle=270}
\caption{Heat conductivity versus system size for $\varepsilon =3.0$.
Statistical and integration errors are less than the symbols.}
\vglue -2.mm
\end{figure}

Conductivities for $\varepsilon = 1.0$ are shown in Fig.~3. Obviously, they
are finite for $T=2$ and $T=8$, showing that the temperature range with
finite conductivities is underestimated in \cite{FK_savin}. For $T=0.5$
and $T=32$ we see a slow increase of $\kappa$ with $N$, over a wide range
of the latter, but it seems to stop for the very largest lattices ($N>1000$).
Finally, for $T=0.125$ and $T=128$ there is a slow increase for all
$N > 300$. The latter could be taken as an indication that $\kappa$ 
diverges for these temperatures, but we think that this would be wrong. 
On the one hand, the increase with $N$ is very slow, much slower than 
in Fig.~4 of \cite{FK_savin}. We would get a similarly fast increase as 
in \cite{FK_savin} if we would use the nominal temperature difference, i.e.
if we would disregard the jumps seen in Fig.~1. On the other hand, the 
data for $0.5 \leq T \le 32$ show us that the saturation of $\kappa$ happens
at larger and larger lattice sizes as we go away from the central energy 
region, just as we have expected. Thus we must expect in any case an 
increase of $\kappa$ for all reachable lattice sizes, and observing 
it does not give any relevant information.

\begin{figure}
\vglue -3.mm
\psfig{file=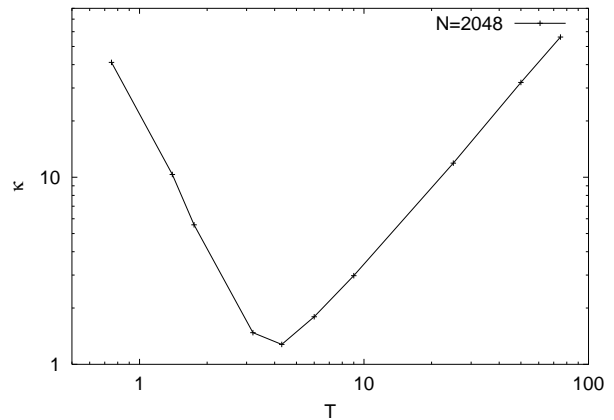,width=5.8cm, angle=270}
\caption{Asymptotic (for $N\to \infty$) heat conductivity versus temperature 
for $\varepsilon =3.0$. The x axis is the temperature, the y axis is conductivity $
\kappa _{{\rm center}}$. Actually, the plotted values of $\kappa$ are those measured 
for $N=2048$, but they seem to be independent of lattice size for $N>1000$.}
\vglue -2.mm
\end{figure}

Analogous results for $\varepsilon =3.0$ are given in Fig.~4. There we 
show data for the temperature range $[0.75,75]$. This is again much larger 
than the range where convergent conductivities were found in \cite{FK_savin}.
This time {\it all} curves become horizontal for large $N$, i.e. the 
conductivity is finite in the entire range. It of course depends strongly 
on $T$, see Fig.~5. It diverges both for $T\to 0$ and for $T\to \infty$, since 
the problem effectively becomes harmonic in both limits.

\begin{figure}
\psfig{file=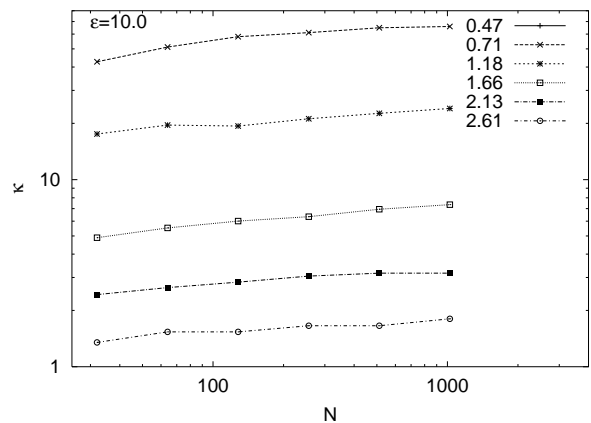,width=5.8cm, angle=270}
\caption{Heat conductivity versus system size for $\varepsilon =10.0$.
Again, statistical and integration errors are less than the symbols.}
\vglue -2.mm
\end{figure}

Finally, our last simulations for the FK model, for $\varepsilon =10$, are 
summarized in Fig.~6. There we only show results for low temperatures, 
$0.5 \le T \le 2.75$. Except for the last temperature, they are all in 
the regime where the authors of \cite{FK_savin} have found divergent $\kappa$.
In contrast, all our curves are either horizontal for all $N$ or become 
horizontal for large $N$, suggesting that $\kappa$ is finite for all finite 
$T$.

\begin{figure}
\psfig{file=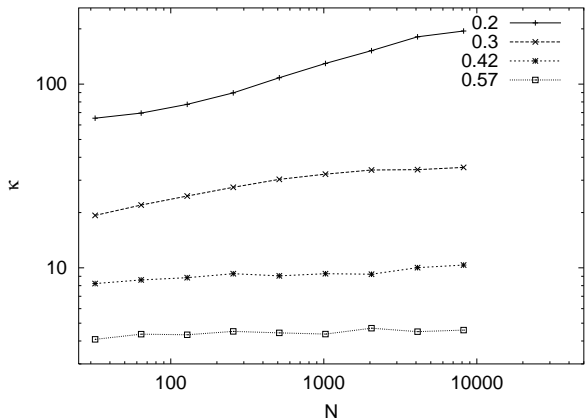,width=5.8cm, angle=270}
\caption{Heat conductivity versus system size for the rotator model.
Again, statistical and integration errors are less than the symbols. }
\vglue -2.mm
\end{figure}

{\bf Rotator model:} For the rotator model, we simulated larger systems, 
with $N$ ranging from 32 to 8192. Conductivities are plotted in Fig.~7 
against $N$ for $T=0.6, 0.45, 0.3,$ and 0.2 (from bottom to top). According 
to \cite{R_savin}, the phase transition from a 
high-$T$ phase with finite conduction to a low-$T$ phase with infinite 
conduction occurs at some $T_c$ between $0.2$ and $0.3$. According to that,
the lowest three curves in Fig.~7 should become flat for $N\to\infty$, while 
the uppermost should continue to grow. This is not what is found, although
our values for $T=0.2$ and 0.3 agree numerically quite well with those 
of \cite{R_savin}. But while the curve for $T=0.2$ increases with the same 
average slope as in Fig.~2 of \cite{R_savin}, it is definitely S-shaped and
stops to rise for the largest values of $N$.

{\bf Conclusion}: In this paper we studied the size dependence of the 
effective finite size conductivity of nonlinear 1D lattices, as a function
of temperature. We used straightforward but high statistics simulations 
to show that there are no indications of the phase transitions suggested 
in \cite{R_savin,R2_savin,FK_savin} on the basis of similar simulations.
For the FK model, this is in agreement with expectations, since phonons 
should have a finite free path in this model for all finite temperatures.
For the rotator model it is less obvious. It suggests that the blocking 
of the propagation of soft phonons by localized excitations \cite{R_savin}
is effective at all finite temperatures. It becomes of course less and 
less important as $T\to 0$, since the density of such excitation decreases 
exponentially with $1/T$. But it is present at all finite $T$, and it becomes 
dominant for $N\to\infty$.

\vspace*{0.3cm}

P. G. wants to thank Roberto Livi and Antonio Politi for very helpful
discussions.


\begin{thebibliography}{30}
\bibitem{foot1}  For dimensions $d\geq 3$, soft phonons should not be
important and the conductivity should be finite. For $d=2$ one expects a
logarithmic divergence with $N$ \cite{lippi}, but simulations \cite{FPU_yg}
show also a power divergence.
\end{thebibliography}
\end{document}